\documentclass[letter]{aa} 

\usepackage{lipsum}  
\usepackage{graphicx}
\usepackage{txfonts}
\usepackage[colorlinks, citecolor=blue]{hyperref}
\usepackage{multirow}

\usepackage{soul}    
\begin{document} 

\newcommand{\Ho}{\mbox{$H_0$}}
\newcommand{\ang}{\mbox{{\rm \AA}}}
\newcommand{\abs}[1]{\left| #1 \right|} 
\newcommand{\kms}{\ensuremath{{\rm km\,s^{-1}}}}
\newcommand{\zabs}{\ensuremath{z_{\rm abs}}}
\newcommand{\zem}{\ensuremath{z_{\rm quasar}}}
\newcommand{\cmsq}{\ensuremath{{\rm cm}^{-2}}}
\newcommand{\ergs}{\ensuremath{{\rm erg\,s^{-1}}}}
\newcommand{\ergsa}{\ensuremath{{\rm erg\,s^{-1}\,{\AA}^{-1}}}}
\newcommand{\ergscm}{\ensuremath{{\rm erg\,s^{-1}\,cm^{-2}}}}
\newcommand{\ergscma}{\ensuremath{{\rm erg\,s^{-1}\,cm^{-2}\,{\AA}^{-1}}}}
\newcommand{\msyr}{\ensuremath{{\rm M_{\rm \odot}\,yr^{-1}}}}
\newcommand{\nhi}{n_{\rm HI}}
\newcommand{\fhi}{\ensuremath{f_{\rm HI}(N,\chi)}}
\newcommand{\refs}{{\bf (refs!)}}
\newcommand{\Av}{\ensuremath{A_V}}
\newcommand{\lya}{Ly$\alpha$}
\newcommand{\Hb}{H-$\beta$}
\newcommand{\OIII}{\ion{O}{iii}}
\newcommand{\OII}{\ion{O}{ii}}
\newcommand{\OI}{\ion{O}{i}}
\newcommand{\HI}{\ion{H}{i}}
\newcommand{\HeII}{\ion{He}{ii}}
\newcommand{\HH}{\ensuremath{{\rm H}_2}}
\newcommand{\qso}{J1259+0309}
\newcommand{\qsolong}{SDSS J125917.31+030922.5}

\newcommand{\fcla}{French-Chilean Laboratory for Astronomy, IRL 3386, CNRS and U. de Chile, Casilla 36-D, Santiago, Chile \label{fcla}}
\newcommand{\iap}{Institut d'Astrophysique de Paris, CNRS-SU, UMR\,7095, 98bis bd Arago, 75014 Paris, France \label{iap}}
\newcommand{\ioffe}{Ioffe Institute, {Polyteknicheskaya 26}, 194021 Saint-Petersburg, Russia \label{ioffe}}
\newcommand{\uchile}{Departamento de Astronom\'ia, Universidad de Chile, Casilla 36-D, Santiago, Chile \label{uchile}}
\newcommand{\eso}{European Southern Observatory, Alonso de C\'{o}rdova 3107, Vitacura, Casilla 19001, Santiago, Chile \label{eso}}
\newcommand{\victoria}{National Research Council Herzberg Astronomy and Astrophysics, 5071 West Saanich Road, Victoria, B.C., Canada, V8Z6M7 \label{victoria}}

   \title{
   The rich galactic environment of a H$_2$-absorption selected quasar\thanks{Based on data collected at the Very Large Telescope from the European Southern Observatory under Prog. IDs 111.24UJ.004 and 105.203L.001}} \subtitle{\lya\ mapping down to the inner kiloparsecs via perfect natural coronagraphy and integral field spectroscopy
   }

   \author{   
%
F. Urbina\inst{\ref{uchile}} 
\and
 P.~Noterdaeme\inst{\ref{iap},\ref{fcla}}
  \and
  T.A.M. Berg\inst{\ref{victoria}}
  \and
   S.~Balashev\inst{\ref{ioffe}}
   \and
   S.~L\'{o}pez\inst{\ref{uchile}}
   \and
   F.~Bian\inst{\ref{eso}}
  }
  
   \institute{\uchile \and \iap \and \fcla \and \victoria \and \ioffe \and \eso}

   \date{}

 
  \abstract
  {We present the first VLT/MUSE observations of a quasar featuring a proximate molecular absorption system, SDSS J125917.31+030922.5.
  The proximate damped \lya\ absorption acts as a natural coronagraph, removing the quasar emission over $\sim$40~{\AA} in wavelength, and allows us to detect extended \lya\ emission without the necessity of subtracting the quasar emission. This natural coronagraph 
  permits the investigation of the quasar environment down to its inner regions ($r < 20$ kpc), where galaxy interactions or feedback processes should have the most noticeable effects. 
  Our observations reveal a dense environment, with a highly asymmetric \lya\ emission within $2"$ ($\sim 15$ kpc), possibly shaped by a companion galaxy, and a southern extension of the nebulae at about 50~kpc, with rotation-like kinematic signature. 
  The width of the \lya\ emission is broadest closer to the quasar, indicating perturbed kinematics as expected if interactions and significant gas flows are present. The foreground absorbing system itself is redshifted by $\approx $400 km/s relative to the background quasar, and therefore is likely arising from gas moving towards the quasar. 
  Finally two additional \lya\ emitters are detected with $>10\,\sigma$ significance at 96 and 223 kpc from the quasar,  making this field overdense relative to other similar observations of quasars at $z\sim 3$. 
  Our results support the hypothesis that quasars with proximate neutral/molecular absorption trace rich environments where galaxy interactions are at play and
   motivates further studies of \HH-selected quasars to shed light on feeding and feedback processes.  

  }

   \maketitle
%

\section{Introduction}

Thanks to their tremendous luminosity, quasars have been used since their discovery in the 1960s as powerful cosmic probes. Their spectra indeed carry a wealth of absorption signatures from the various \textit{intervening} gaseous environments encountered along the line of sight to the observer, ranging from the 
diffuse ionized intergalactic medium \citep[e.g.,][]{1998Rauch} to large column densities of neutral gas that  imprint characteristic damped Lyman-alpha systems \citep[DLAs, with $N(\HI) \geq 2\times 10^{20}$\cmsq, see][]{2005_DLA_wolfe}.

\textit{Associated} systems are believed to originate within the quasar environment.
Due to their physical proximity, a low velocity difference between the quasar and the associated system is expected, dominated by the kinematics of the gas rather than the Hubble flow. This motivated the first systematic studies of associated systems by selecting samples of \textit{proximate} DLAs, defined as systems with velocity differences $|\Delta\rm v|< 5000$ km s$^{-1}$ \citep[PDLAs, see][]{2010Ellison}. The detectability of such systems in \HI\ depends on the competition between clustering in quasar environment and ionization by the intense radiation field of the later \citep{2008Prochaska}.

More recently, \citet{2019Noterdaeme} uncovered a population of proximate molecular absorption systems. Because the presence of H$_2$ requires a shielding \HI\ layer \citep[see e.g.][]{2014Sternberg}, proximate H$_2$ absorbers are expected to be PDLAs. The incidence rate of proximate H$_2$ absorbers was found to be an order of magnitude higher relative to the intervening H$_2 $ systems, indicating that proximate H$_2$ systems are indeed tracing environments surrounding the quasars. The velocity of the foreground absorber relative to the background quasar is likely due to the kinematics of the gas, where positive and negative velocity differences suggest accretion and outflow processes, respectively, in line with an observed trend between this velocity and the gas metallicity \citep{2023Noterdaeme}. Detailed analysis of optical and millimetre data then permits further investigation of outflowing \citep{2021aNoterdaeme} or merging (Balashev et al. in prep) processes in these systems.

\begin{figure*}
    \centering
    \includegraphics[width=0.87\hsize]{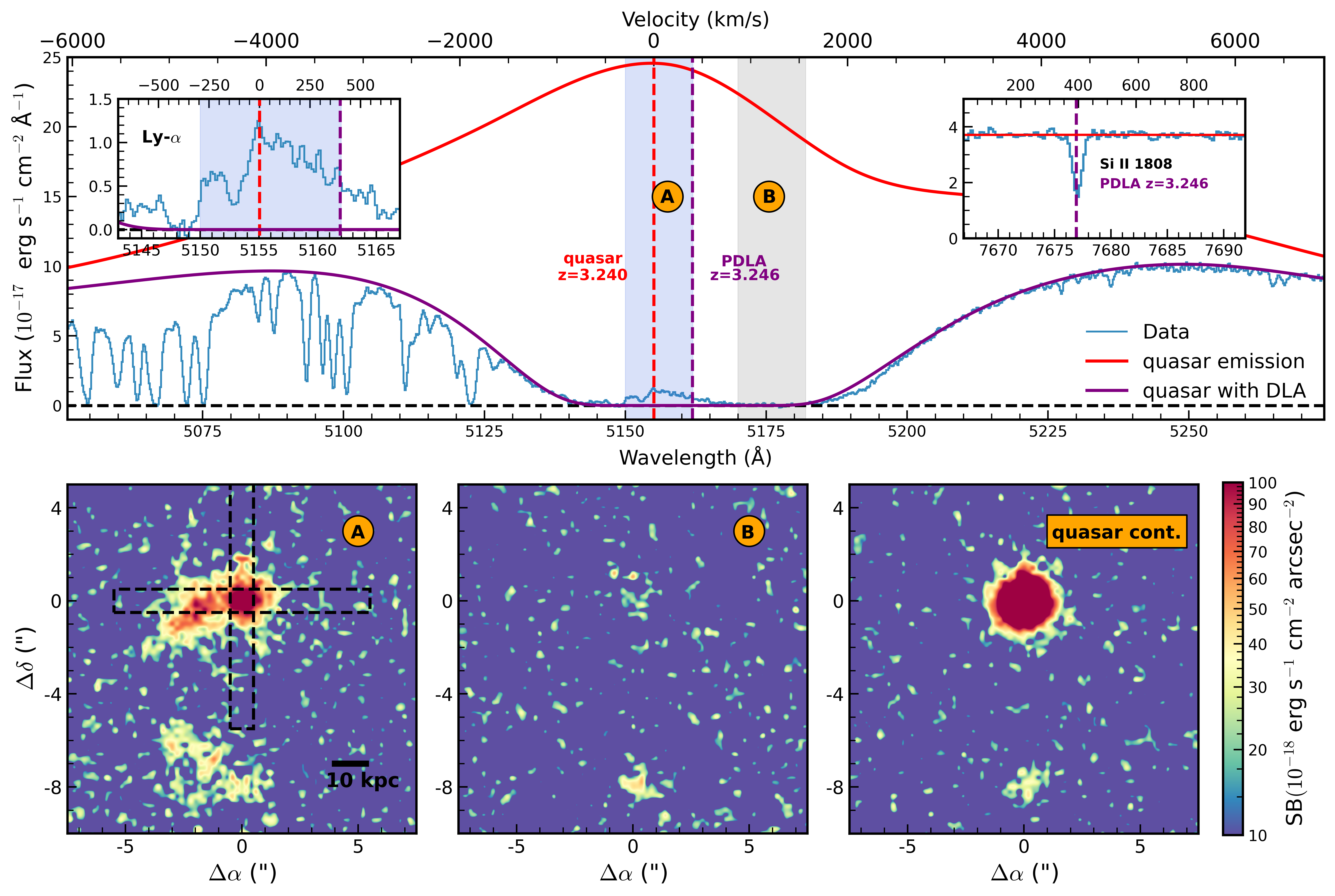}
    \caption{\textbf{Top row:} J1259+0309 X-shooter 1D combined spectrum 
    around systemic \lya\ wavelength.  The red solid line shows the reconstructed quasar emission to which the PDLA Voigt profile model ($\log N(\HI)\approx21.4$) is then applied to obtain the solid purple line. Vertical red and purple dashed lines indicate the expected location of lines at the quasar systemic redshift from CO(3-2) emission (also the zero-point in the velocity from the top axis) and the PDLA, respectively. The inner sub-panels present zoom in around 
    wavelength regions containing the \lya\ emission in the PDLA core and a low-ionisation absorption line. 
        \textbf{Bottom row:} MUSE narrow-band images from the wavelength regions A and B indicated in the top panel by blue and grey vertical stripes, respectively. The third panel shows an only quasar continuum region at 5400-5412$\AA$. All images were smoothed using a Gaussian kernel, share the same scale and coordinates are relative to the quasar central position in the sky. In panel A, horizontal dashed rectangles illustrate the position the two X-shooter slits employed to obtain the combined 1D spectrum. A scale of 10 proper kpc ($\approx 1.3 "$) is shown as a black horizontal solid line.
    \label{fig:xshooter NB images}}
\end{figure*}

The above observations have not yet been connected to the quasar emission properties. Integral field spectroscopy (IFS) has proved to be invaluable in providing 3D maps of the emission in quasar fields, even reaching an striking 100\% detection rate of extended \lya\ emission around bright $z \approx 3$ quasars, with scales reaching up to hundreds of kpc \citep{2016Borisova}. 
%
%
%
However, the outshining continuum and \lya\ emission from the central engine, even if expected to arise from a tiny region, poses a severe challenge to investigate structures at galactic scales ($\sim 10$~kpc), closer to the quasar itself, where outflows, inflows and galaxy interactions are expected to play a crucial role.
Indeed, regardless of the consistent detection, the ability to recover the resolved emission depends on the efficiency of quasar point spread function (PSF) 
subtraction methods.  
Empirical PSF methods generally perform well at large distances from the central quasar but leave large residuals in the inner regions \citep[see examples in][]{2019qso_museum, 2019RequiemSurvey}. Other methodologies involve modeling the quasar spectra and then subtracting the model for each spaxel of the data cube \citep{2017NorthPDLA}, but the accuracy of the extracted emission strongly depends on the modeling, which may also introduce biases by involving the use of quasar template spectra. 

In this letter, we present the first IFS observations of a proximate H$_2$ system, in the 
quasar SDSS J125917.31+030922.5 (hereafter J1259+0309). A natural coronagraph is provided by the corresponding high-column density PDLA, enabling \lya\ nebula extraction
without the necessity of modelling and subtracting the quasar light, making it the most precise mapping of nebular emission at low projected distances to date.
This allows us to trace the gas structure in a complex, likely interacting environment. 

\begin{figure*}
    \centering
    \includegraphics[width=0.9\hsize]{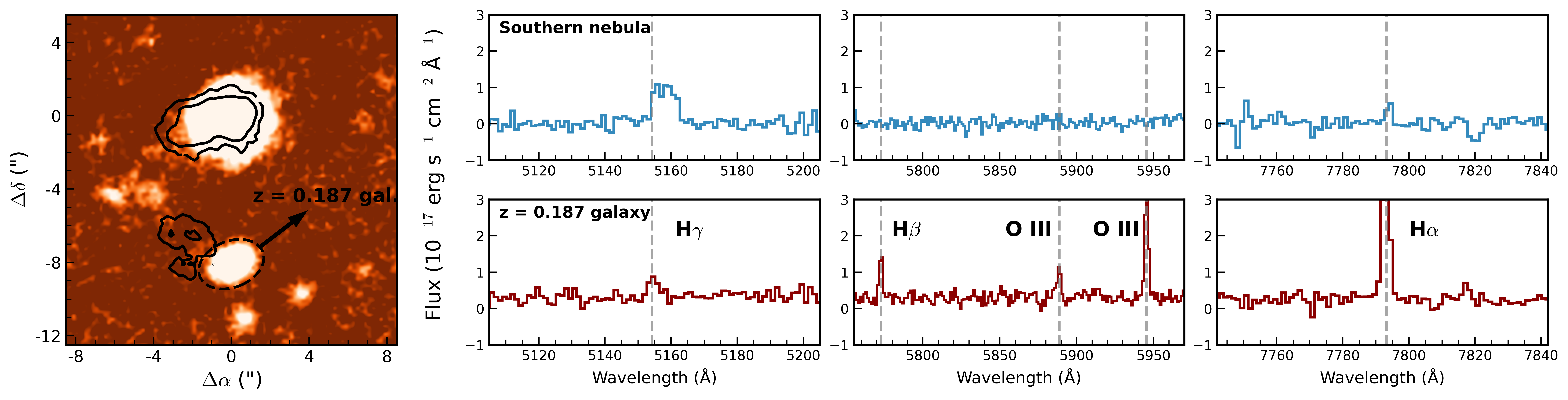}
    \caption{\textbf{Left-most panel:} White light image of the MUSE data cube around the quasar (at (0,0) position) prior to subtracting contamination sources. 
    5 and 10\,$\sigma$ contours of the extended \lya\ nebula are overlaid. Enclosed by a dashed elliptical contour is  the $z_{\rm gal}=0.187$ line emitting galaxy.  
    \textbf{Other panels:} Spectra of the southern portion of the nebula (top row) and the line emitting galaxy (bottom row) shown in three different wavelength portions covering H$\alpha$, 
    H$\beta$+[\OIII] and H$\gamma$ at $z_{\rm gal}$, the latter coinciding with Ly$\alpha$ at $z_{\rm quasar}=3.24$. 
    }
    \label{fig:white light}
\end{figure*}
\section{Observations}

IFS observations of J1259+0309 were carried out by the Multi Unit Spectroscopic Explorer (MUSE) instrument (\textcolor{blue}{Bacon et al. 2010}), mounted on UT4 at the Very Large Telescope (VLT). Observations were executed 
under good seeing conditions (average $0.7\arcsec$). 
Six 10 minutes long exposures were taken with 
different position angles (0$^\circ$, 22.5$^\circ$, 45$^\circ$, 90$^\circ$, 112.5$^\circ$ and 135$^\circ$), for better spatial sampling.  


Raw exposures were reduced using 
ESO MUSE Data Reduction Software \citep[pipeline version 2.8.9;][]{2020MUSE_reduction}. We ran two iterations of the \textit{scipost} recipe, the first to obtain data images for producing a sky mask to be used in the second iteration. 
Finally, the Zurich Atmosphere Purge  \citep{2016zap} software was used to remove any residual pattern from the sky subtraction procedure.
To account for correlated noise, the final variance cube is scaled to match the measured variance of the background. The PSF's full width at half maximum (FWHM) is measured to be 0.7" in processed datacube.

In this paper, we also use spectroscopic data from VLT/X-shooter 
and adopt the systemic redshift of the quasar host galaxy as $z_{\rm syst}=3.2405 \pm 0.0011$ based on CO(3-2) emission from NOEMA observations  \citep{2023Noterdaeme}. We use $z_{\rm syst}$ as the reference for zero velocity in the following sections. This $z_{\rm syst}$ implies a spatial scale of 7.5 pkpc per arcsec for a flat $\Lambda$CDM  cosmology with $\Omega_m = 0.3$ and $H_0 = 70$~km\,s$^{-1}$\,Mpc$^{-1}$.



\section{Methodology and results}
\label{sec: methods}

\begin{figure*}
    \centering
    \includegraphics[scale=0.57]{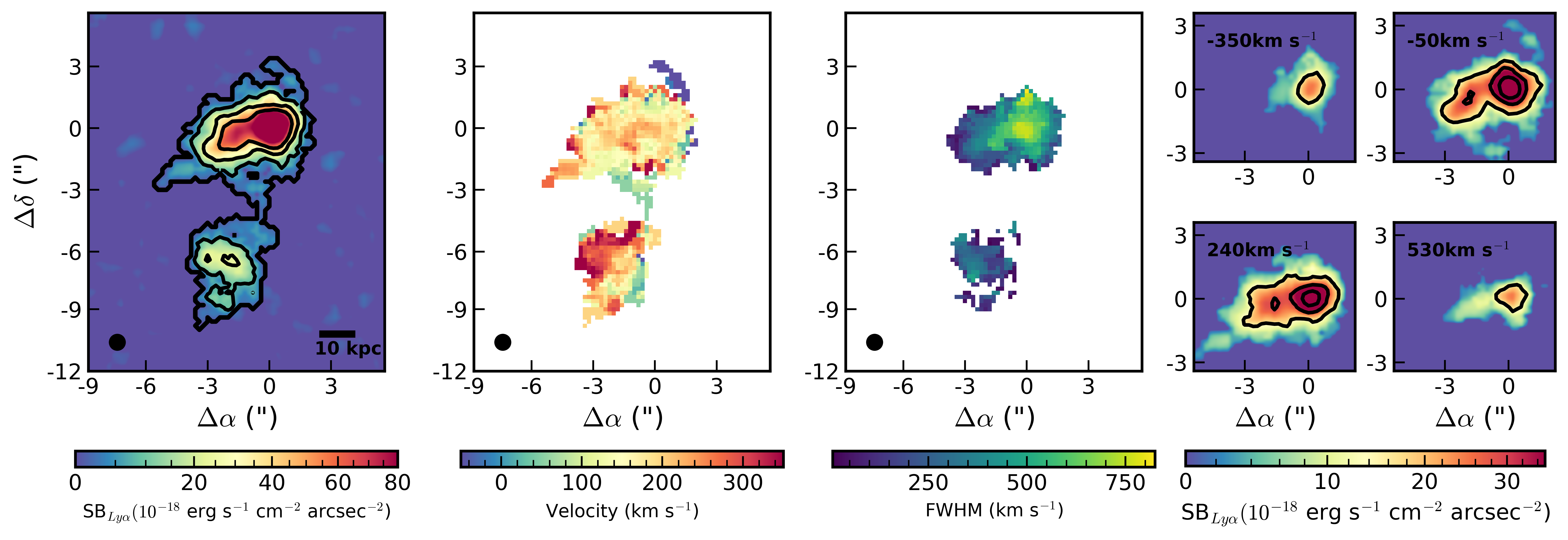}
    \caption{\textbf{First panel:} 
    Optimally extracted \lya\ nebula (zeroth moment) from continuum subtracted MUSE datacube of J1259+0309. The spatial projection of the segmentation mask is shown as a black thick contour around the detected emission and it is approximately a 2.5$\sigma$~=  2.45 $\times 10^{-18}$ erg s$^{-1}$cm$^{-2}$ arcsec$^{-2}$ contour. Thin black contours enclose 5, 10 and 15 $\sigma$ regions. The black horizontal line indicates a scale of 10 pkpc.  For reference, the PSF is depicted by a black circle with diameter of 0.7", i.e., the PSF FWHM. 
    \textbf{Second panel} Velocity map of the emission with zero point defined by $z_{sys}$.
    \textbf{Third panel:}
    Second moment map to show the equivalent Gaussian FWHM, where the instrumental FWHM$_{ins} \approx 170$\,\kms\ has been subtracted in quadrature. Each image has a linear projected size of 14"$\times$17" and coordinates are relative to the quasar position.
    \textbf{Last panel:} Zoom-ins of the inner region of the emission in four $\approx$ 290\,\kms\ wide velocity bins with central value indicated in the top left corner of each image. 
    %
    Isophotal contours of surface brightness values of 15, 30 and 45 $\times 10^{-18}$ erg s$^{-1}$ cm$^{-2}$ arcsec$^{-2}$ are displayed. 
    \label{fig:ly-alpha nebula}}
\end{figure*}

    \subsection{The PDLA as perfect natural coronagraph}

    Observing \lya\ emission in quasar fields normally requires precise modelling of the quasar PSF as it heavily dominates the 
    emission. For IFS data, this is generally done empirically by estimating the PSF from broad band images extracted from the same data cube (see \cite{2016Borisova} or \cite{2019RequiemSurvey} for examples). However, the outshining quasar emission itself inevitably leaves residuals that prevent measuring faint nebular \lya\ emission in the inner regions, impeding any characterisation of the emission at the scale of the quasar host or nearby companion galaxies.     
    For J1259+0309, the presence of a PDLA ($z_{\text{abs}}~=~3.2461\pm~0.0001$) with large \HI\ column density ($\log N(\HI)\approx21.4$) efficiently suppresses the quasar light over a wide wavelength range (optical depth higher than 5 over $\approx40\AA$) so, for the first time, there is no need to perform neither PSF nor continuum subtraction and  we can directly extract the extended \lya\ emission, including the relatively narrow component at the core of the DLA profile 
   (see Fig. \ref{fig:xshooter NB images}). 
    In short, the PDLA acts as a natural 
    coronagraph that suppresses the point source emission well before any 
    dispersive process due to the Earth's atmosphere, the telescope or the instrument.
    
   

\subsection{Identification and subtraction of a low redshift galaxy}

 The left panel of Fig.~\ref{fig:white light} shows a white light image of the data cube, in which several sources are detected. Within this $16" \times 16"$ region, we identified only one additional line emitting source, which turns out to be a $z=0.187$ galaxy and whose H$\gamma$ emission, falling at $5153 \AA$, could in principle contaminate the high-$z$ \lya\ map at this position. 
We predicted the H$\gamma$ emission based on the strength of the H$\alpha$ and H$\beta$ lines, assuming recombination and values reported in Table B.7 from \cite{2003difuse_u}, and subtracted the emission from the cube on a spaxel-by-spaxel basis wherever we managed to detect H$\alpha$ with S/N $\geq 3$, despite the strength of the H$\gamma$ emission being below the noise level for most subtracted spaxels.

\subsection{Optimal extraction of the \lya\ nebula}
    We employed similar methods as outlined in \cite{2016Borisova}  and \cite{2019RequiemSurvey}, in which the only pre-detection process is background subtraction at spaxels where the quasar is not detected. A S/N cube is then created from the results of the last step, which is converted into a binary cube, where unit values indicate voxels with S/N $\geq 2.5$ (zero otherwise). Finally, a 3D segmentation mask is obtained by running a friends-of-friends algorithm on this binary cube. The largest group of connected voxels is considered as the final mask. An extended explanation of the methodology will be presented in Urbina et al. (in prep.). 
    The optimally extracted \lya\ nebula (zeroth moment) and its first and second moments are shown in Fig. \ref{fig:ly-alpha nebula}.
%
%
In addition to the main emission around the position of the quasar, we detect an elongated secondary emission centered $\approx$2~arcsec ($\approx$15 kpc) to the East and a weaker Southern emission connected through a filamentary structure (S/N $\approx 2$). We also present four zoomed in velocity-binned images of the nebula in the inner regions for better visualization of the structures 
at those scales.

\subsection{Additional \lya\ emitters in the field}

\begin{table}
		\centering
		\caption{Properties of LAE candidates in the field of \qso. 
  }
		\begin{tabular}{ccc}
			\hline \hline
			&{LAE 1} &  {LAE 2}  \\
			\hline
        RA (J2000) &   12:59:16.777 & 12:59:15.922  \\
		DEC (J2000)  &  +03:09:34.57    &      +03:09:45.97 \\
  Impact parameter (pkpc)&  96 & 223 \\
            Redshift & 3.2455& 3.2457 \\		
            Integrated flux & \multirow{2}{*}{3.8 $\pm 1.1$} & \multirow{2}{*}{4.3$\pm 1.1$}\\
            {\small (10$^{-17}$ erg s$^{-1}$ cm$^{-2}$ )} &  & \\
            Peak \lya\ flux &     \multirow{2}{*}{8.0$\pm 1.0$}   & \multirow{2}{*}{9.1$\pm 1.1$}  \\
             {\small (10$^{-18}$ erg s$^{-1}$ cm$^{-2}$ \AA$^{-1}$)} & & \\
            Peak significance ($\sigma$)& 11.5 & 11.4\\
			\hline
		\end{tabular}
		\label{tab: LAES}
	\end{table}
 \begin{figure}[h]
    \centering
    \includegraphics[width=1\hsize]{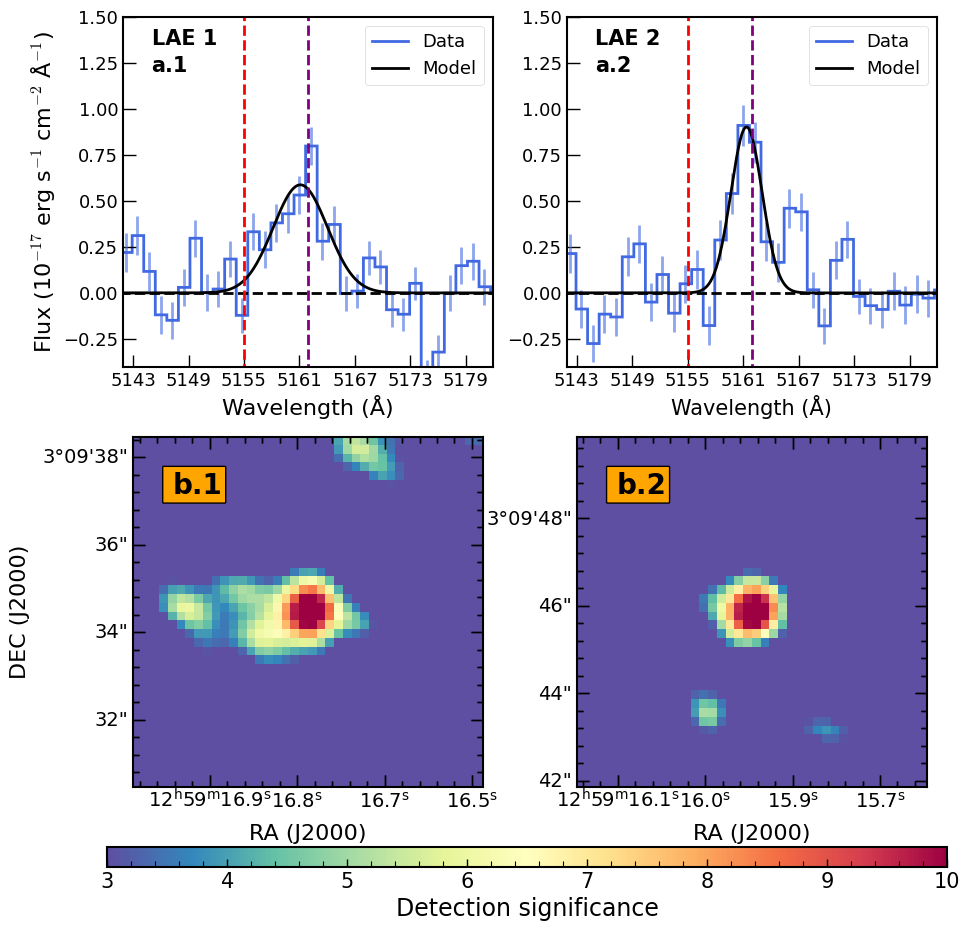}
    \caption{\textbf{(a.1, a.2):} LAE candidates spectra obtained obtained using 1" apertures centered at the coordinates reported in Table \ref{tab: LAES}. Red and purple dashed lines show the expected wavelength of the emission at the quasar systemic and PDLA redshift.  Black solid line shows a gaussian model fitted to the data. Propagated $1\sigma$ errors are shown as vertical blue lines.  
    \textbf{(b.1, b.2):}
    LSDCat output detection significance maps at the wavelength channel where the peak of \lya\ emission is reached.}
    \label{fig:laeS}
\end{figure}
To explore the wider environment around \qso, we searched for potential \lya\ emitters (LAE) in the cube using the LSDCat weak line emitting source detection software \citep{lsdcat}. 
LSDCat employs a matched filtering approach which is applied in spatial and then in the spectral direction respectively. For J1259+0309, we used a 2D Moffat profile with FWHM $\approx$ 0.7" in the spatial direction and a gaussian profile with FWHM of 300~\kms\ for the spectral direction, i.e., a compact source with a narrow line emission. The output of the software is detection significance cube. 
In Fig.~\ref{fig:laeS}, we show the extracted spectrum and narrow-band images of the two LAE candidates detected 
this way 
and summarize their properties in Table~\ref{tab: LAES}.


\begin{figure}[h]
    \centering
    \includegraphics[scale=0.6]{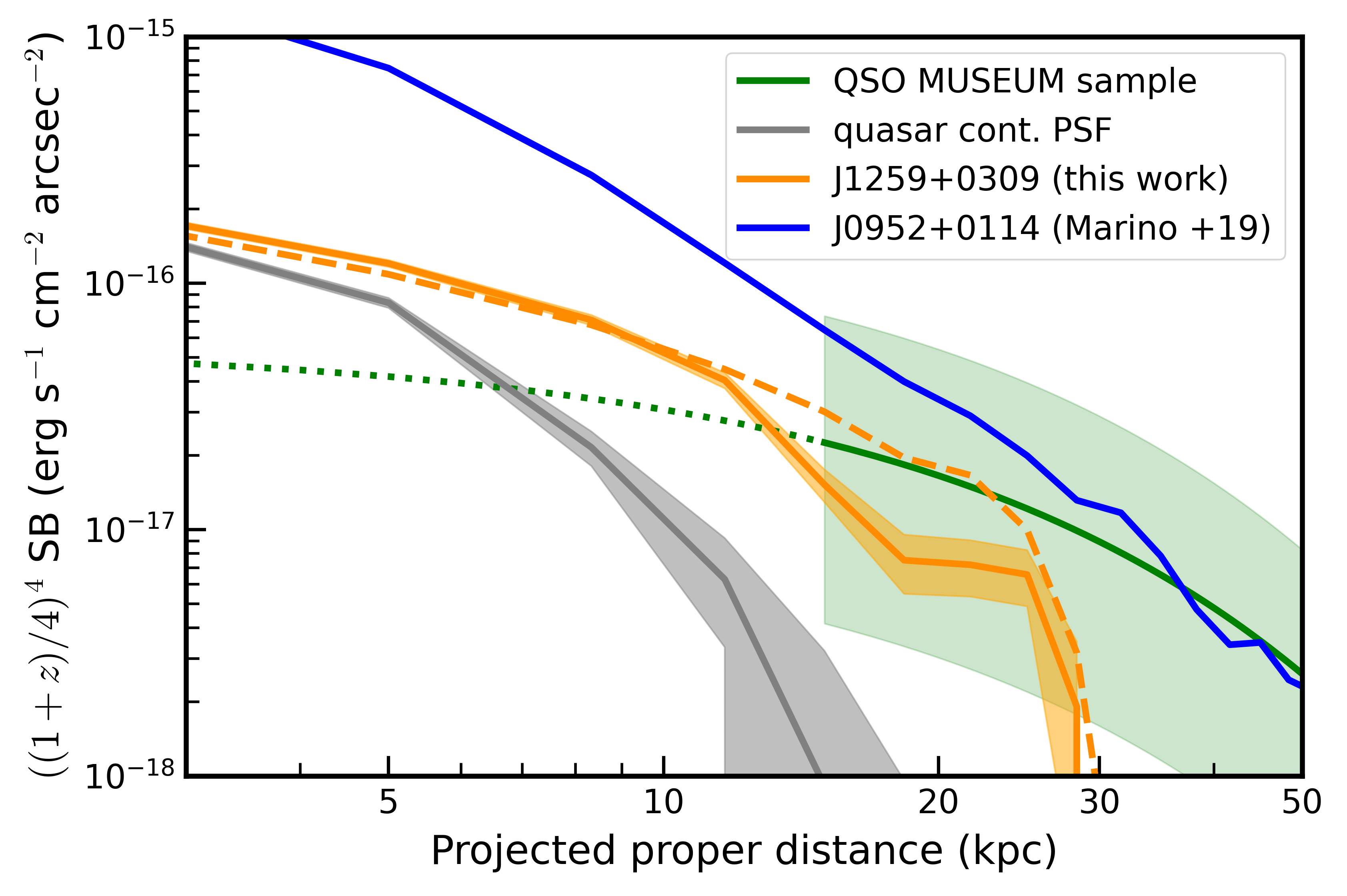}
    \caption{Circularly average \lya\ surface brightness profiles (corrected for redshift dimming) as a function of projected distance from the central quasar. Orange, blue and green solid lines stand for \qso\ (this work, with 1$\sigma$ error as shaded area), J0952+0114 \citep{2019Marino} and the average of the QSO MUSEUM sample \citep[][with 1$\sigma$ scatter as light green area]{2019qso_museum}, respectively. The green dotted line is an extrapolation of the latter 
    towards low radius. 
    The orange dashed line exhibits the eastward average profile for \qso\ (see Sec. \ref{sec: discussions}). The sharp drop at 30 kpc in \qso\ profile is due to sensitivity limitations. 
    The PSF shape, derived from the quasar continuum, is shown as a grey solid line (with 1$\sigma$ error) and scaled 
    for visual purposes. 
    }
    \label{fig:SB profile}
\end{figure}

\section{Discussion}
\label{sec: discussions}

In Fig.~\ref{fig:SB profile}, we present the circularly averaged surface brightness (SB) profile of the \lya\ emission around \qso, together with that of SDSS\,J095253.83+011421.9 \citep[hereafter J0952+0114][]{2019Marino} and the best exponential fit to the QSO museum sample \citep{2019qso_museum}. Apart from a few noticeable cases, most \lya\ emission in this sample is concentrated within 50 kpc from the quasar, and has a relatively isotropic profile with a median minor-to-major axis ratio of 0.73. 
%
However, PSF contamination impedes constraining the \lya\ SB profile within the inner 2-3”.
This is unfortunate since the close environment of quasars (within a few 10 kpc) is where outflowing or infalling gas and galaxy interactions should have most noticeable effects and hence key to 
understanding the baryonic cycle or galaxy interactions. Moreover, precisely determining the SB profile of the inner regions can provide some insights into the powering mechanism of the nebulae. Post-processing cosmological simulations by \cite{2022Costa} suggest an approximately flat profile at low projected distances is not possible to recover when only assuming recombination and collisional excitation. As the profile shown in Fig. \ref{fig:SB profile} is indeed approximately flat up to $\approx 10$ kpc, we may conclude that  mechanisms such as resonant scattering  can be relevant in this case.

MUSE observations of J0952+0114, where PSF subtraction is not required due to the presence of a PDLA\footnote{In J0952+01114 there can still be quasar continuum contamination as zero flux levels are not reached in the \lya\ portion of the spectrum.}, show a much higher flux in the central regions compared to what is expected from extrapolating the MUSEUM profile but did not reveal much structure either. This particular quasar originally attracted attention because it lacks the broad \lya\ component expected from the broad metal line emission \citep{2004Hall}. \citet{2016Jiang} later showed the presence of a PDLA which covers only the continuum and the broad-line region but is filled with relatively narrow ($\sim 1000$~\kms) emission, explaining the difficulty in identifying the PDLA as such.
This PDLA exhibits strong absorption from silicon in its excited fine-structure state (Si\,{\sc ii}*), which may indicate high compactness of the associated gas \citep{2020Fathivavsari_compactness}. Additionally, the quasar spectrum shows a higher line emission-to-continuum ratio compared to quasar composite spectra, suggesting differential reddening, where dusty gas in the PDLA does not fully cover the metal emission regions either. Regardless of its exact origin, the extended \lya\  emission in J0952+0114 may still be dominated by a central point-like source that is observationally smoothed by the PSF over a few arcseconds. This would explain the very strong Ly-alpha emission in the DLA as well as the smooth symmetric shape centered on the quasar position, with a lack internal structure at $r < 20$ kpc ($\sim 3"$).

In the case of \qso, the central point source emission is efficiently suppressed over a wide wavelength range and the remaining emission has a surface brightness closer to expected from extrapolating the peripheral emission profile. While we do not know the actual extent of the absorbing gas, our \lya\ mapping is not blinded by the central emission, which allows us to resolve structures presented in Fig.~\ref{fig:ly-alpha nebula}. 
We indeed observe a clear asymmetric shape of \lya\ emission even at small separations of $2"$ ($\approx 15$ kpc) from the quasar, which extends eastwards. For the inner regions of the nebula (without the southern extension) we measured a flux weighted minor-to-major axis ratio of 0.48. Only $\approx10$\% of \lya\ nebulae have a minor-to-major axis ratios lower than \qso\ \citep{2019qso_museum}.

The relative velocity of the PDLA with respect to $z_{\rm syst}$ is $\Delta v = 430 \pm 80 $ km/s -- the PDLA is at a higher redshift 
than the quasar -- 
which suggests 
that absorbing gas must be moving towards the quasar.
The gas traced by the DLA could hence be tracing a feeding stream or to another galaxy in the group that may eventually merge with the quasar host. 
This system is also a low-metallicity proximate H$_2$ absorber \citep[$Z\approx 0.03\,Z_{\odot}$, $\log N$ (H$_2) \approx 19.10$;][]{2023Noterdaeme} with little to no dust reddening. Therefore, for the \HI-H$_2$ transition to occur under the influence of the quasar radiation, we anticipate a relatively high volume density \citep{2019Noterdaeme}\footnote{{The excitation of H$_2$ suggests the cloud is located at a few 10~kpc, i.e. outside the quasar host galaxy, with density $n > 10^3$\,cm$^{-3}$. This also means that most of the Ly$\alpha$ photons from the quasars are actually emitted within a parsec to sub-parsec size region to be covered by such a compact cloud. This is similar to the size of the broad line region typically estimated using other quasar emission lines.}}.
%
We finally note that the quasar is also a broad absorption line (BAL) quasar, which suggest quasar-driven outflows may be present as well.  Nevertheless, this is not necessarily related to the PDLA since less than 10\% of quasars are expected to be BALs \citep{2006trump}. 

These PDLA properties suggest the quasar is located in a dense environment with ongoing  galaxy interactions.
The \lya\ emission shows consistent signatures of this. 
A positive gradient in FWHM towards the central quasar (see third panel in 
Fig.~\ref{fig:ly-alpha nebula}) suggests disturbed kinematics 
occurring in regions with outflows or infalling material, although the \lya\ emission can be further broadened by radiative-transfer effects only \citep[e.g.][]{2011Laursen}. 
The inner nebula shows enhanced secondary emission to the East, possibly due to a galaxy companion\footnote{Using Subaru/HSC $r$-band imaging we constrained the absolute magnitude of the companion galaxy to be $M_{\rm UV}>-20.7$.}, which drives the asymmetric shape as well.
Additionally, the nebula extends southwards from the central quasar at $7$" ($\sim 50$ kpc). This region shows a different kinematic signature compared to the inner regions, with a velocity gradient may indicate a rotating gas (Fig. \ref{fig:ly-alpha nebula}). Albeit detected only with S/N$\approx2$, a filamentary structure may connect this emission to the material in the immediate vicinity of the quasar.  We also note that the number of LAE candidates detected within the \qso\ field (2 LAE) is $\approx4$ times larger than expected for similar quasar fields at $z\sim3$ (assuming a 2$\sigma$ detection limit of $10^{-18}$ erg s$^{-1}$ cm$^{-2}$ arcsec$^{-2}$ and after completeness corrections, \citealt{2019qso_museum}), suggesting a denser-than-average environment.

The rich environment unveiled by the \lya\ emission around \qso\ suggests that galaxy interactions are actively occurring, on scales of $\sim$ 15 kpc up to $\sim$ 50 kpc. Such a dense environment is not typical for quasars at these redshifts. This suggest that, by selecting the system based on H$_2$ absorption, we have selected an environment richer than average.  As large programs surveying the typical $z\sim 3$ population of quasars are already available, the next step 
is to explore a large number of quasar fields with proximate absorbers (Urbina, in prep) to assess whether selecting quasars with proximate absorbers dictate their environmental properties or the properties seen in \qso\ field can be attributed to the cosmic variance.



\begin{acknowledgements}
F.U. acknowledges support by Subdirection of Human Capital ANID (national MSc 2023/22231861). S.B. is supported by RSF grant 23-12-00166. S.L. acknowledges support by FONDECYT grant 1231187. Authors thank the referee for all comments and suggestions provided.
\end{acknowledgements}

\bibliographystyle{aa} 
\bibliography{article_bib} 

\clearpage
\onecolumn

\end{document}